\documentclass[journal]{IEEEtran}
\usepackage{amsmath}
\usepackage{amssymb}
\usepackage{amsthm}
\usepackage{cite}
\usepackage{color}
\usepackage{xcolor}
\usepackage{caption}
\usepackage{epsfig,latexsym}
\usepackage{float}
\usepackage{fancyhdr}
\usepackage{graphicx}
\usepackage{indentfirst}
\usepackage{mdwmath}
\usepackage{mdwtab}
\usepackage{subfigure}
\usepackage{setspace}
\usepackage{times}
\usepackage{url}
\usepackage{multirow}

\newtheorem{remark}{Remark}
\newtheorem{theorem}{Theorem}

\newtheorem{lemma}{Lemma}

\newtheorem{corollary}{Corollary}

\begin{document}

\title{\Huge{ A Joint Design for STAR-RIS enhanced NOMA-CoMP Networks: A Simultaneous-Signal-Enhancement-and-Cancellation-based (SSECB) Design}}
\author{Tianwei Hou,~\IEEEmembership{Member,~IEEE,}
        Jun Wang,
        Yuanwei Liu,~\IEEEmembership{Senior Member,~IEEE,}
        Xin Sun, \\
        Anna Li and Bo Ai,~\IEEEmembership{Senior Member,~IEEE,}

\thanks{This work is supported by the Fundamental Research Funds for the Central Universities Foundation of China under Grant KWRC21005532. (Corresponding author: Anna Li.)}
\thanks{T. Hou, J. Wang and X. Sun are with the School of Electronic and Information Engineering, Beijing Jiaotong University, Beijing 100044, China (\{email: twhou, wangjun1, xsun\}@bjtu.edu.cn).}
\thanks{B. Ai is with the State Key Laboratory of Rail Traffic Control and Safety, Beijing Jiaotong University, Beijing 100044, China (email: boai@bjtu.edu.cn).}
\thanks{Y. Liu and Anna Li are with School of Electronic Engineering and Computer Science, Queen Mary University of London, London E1 4NS, U.K. (e-mail: yuanwei.liu@qmul.ac.uk, A.li@qmul.ac.uk).}
}

\maketitle
\vspace{-0.1in}
\begin{abstract}
In this correspondence, a novel simultaneous transmitting and reflecting (STAR) reconfigurable intelligent surfaces (RISs) design is proposed in a non-orthogonal multiple access (NOMA) enhanced coordinated multi-point transmission (CoMP) network. Based on the insights of signal-enhancement-based (SEB) and signal-cancellation-based (SCB) designs, we propose a novel simultaneous-signal-enhancement-and-cancellation-based (SSECB) design, where the inter-cell interferences and desired signals can be simultaneously eliminated and boosted. Our simulation results demonstrate that: i) the inter-cell interference can be perfectly eliminated, and the desired signals can be enhanced simultaneously with the aid of a large number of RIS elements; ii) the proposed SSECB design is capable of outperforming the conventional SEB and SCB designs.
\end{abstract}


\section{Introduction}

Reconfigurable intelligent surfaces (RISs) stand as a potential solution of spectral efficiency (SE) enhancement and coverage enhancement for the sixth-generation (6G) wireless networks~\cite{Renzo_two_scenarios_large_zongshu,LIS_magazine_multi-scenarios}.
The received signals can be positively boosted or mitigated by appropriately controlling both the phase shift and the amplitude of the RIS elements. Recently, researchers have shown an increased interest in a novel concept of simultaneous transmitting and reflecting RISs (STAR-RISs), which consider both the transmission and reflection of the RIS elements. The discipline of STAR-RISs is the concept that the incident wireless signals can be reflected within the half-space at the same side of the RIS, whereas they can also be transmitted into the other side of the RIS~\cite{Yuanwei_STAR_RIS_mag,Jiaqi_STAR_RIS_lett}.

In practice, the coordinated multi-point transmission (CoMP) technique is adopted in the third generation partnership project (3GPP) for cellular networks, where the SE and energy efficiency (EE) is limited by the impacts of inter-cell interferences~\cite{3GPP_COMP}. The scarce of SE and EE in 6G has stimulated the development of advanced multiple access techniques, i.e., non-orthogonal multiple access (NOMA)~\cite{NOMA_mag_DingLiu}. One of the main direction of the NOMA-CoMP networks, namely NOMA-coordinated-beamforming (CB), received considerable attention~\cite{NOMA_COMP_Secnarios}. In NOMA-CB, the cell-edge user (CEU) and cell-center user (CCU) are served by its desired BS, where zero-forcing based design was proposed for mitigating the intra-cell and inter-cell interferences by sacrificing the spatial diversity gain of the desired links~\cite{NOMA_COMP_lett,Wan_RIS}, which motivates us to employ RIS for the NOMA-CoMP networks.

By considering the application of the reflected signals at the RIS elements, a pair of main directions of the RIS enhanced networks were proposed, namely signal-enhancement-based (SEB) and signal-cancellation-based (SCB) designs~\cite{RIS_our_survey}. \textcolor[rgb]{0.00,0.00,1.00}{ By assuming that multiple waves are co-phased at the receivers, the received signal can be significantly boosted~\cite{hou2020reconfigurable}. However, both desired signals and interferences increase with the aid of SEB designs, where the rate ceiling occurs in the high-SNR regime. In contrast, aligning the reflected signals for signal cancellation is another way for implementing RISs~\cite{hou_MIMO_NOMA_SCB}, where the performance can be enhanced in the high-SNR regime, but the enhancement of the desired signal is ignored. The above phenomenons motivate us to simultaneously enhance and mitigate the desired signals and the interferences, respectively.} \textcolor[rgb]{0.00,0.00,1.00}{However, the research to date has tended to focus on the SEB-only or SCB-only designs rather than simultaneous-signal-enhancement-and-cancellation-based (SSECB) designs, which is capable of beneficially enhancing the network performance in both the medium-low-SNR and high-SNR regimes.} Fortunately, with the aid of the STAR-RIS, the SSECB designs become possible for implementing NOMA-CoMP networks.

However, the research of STAR-RIS enhanced NOMA-CoMP networks is technically challenging. For example, on the one hand, the passive beamforming design of STAR-RIS for the SSECB design is still an open issue. On the other hand, the minimal required number of RIS elements is also worth estimating. Hence, in order to provide the benchmark scheme of the STAR-RIS enhanced NOMA-CoMP networks, the main contributions can be summarized as follows:
\begin{itemize}
  \item We first propose a novel RIS enhanced CoMP network by NOMA-CB, where each BS serves its own users and RIS array.
   By assuming that each STAR-RIS elements are capable of both transmitting and reflecting the incident signals, the signals can be controlled in full-spaces.
  \item We then propose a new SSECB design in the STAR-RIS enhanced NOMA-CoMP networks, where the signal and interference can be boosted and eliminated by the reflection and the transmission of the STAR-RIS elements for enhancing the network's performance. The minimal required number of RIS elements is derived in the case of strong-line-of-sight (LoS) links.
  \item Our simulation results illustrate that 1) The minimal required number of RIS elements is mainly impacted by the distances and path-loss exponent of the interference and transmission links; 2) Since the small-scale fading gains are random variables, the inter-cell interference can be cancelled perfectly in the case that the number of RIS elements is large enough; 2) The proposed SSECB design is capable of outperforming the conventional SEB and SCB designs.
\end{itemize}

\section{System Model}

\begin{figure}[t!]
\centering
\includegraphics[width =3 in]{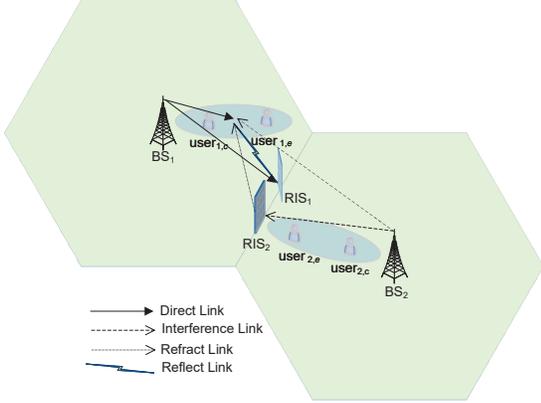}
\caption{System model of the STAR-RIS enhanced NOMA-CoMP networks.}
\label{system_model}
\end{figure}

The system model of the STAR-RIS enhanced NOMA-CoMP networks is shown in Fig.~\ref{system_model}, where two cells are deployed for serving the CCUs and CEUs simultaneously in downlink. \textcolor[rgb]{0.00,0.00,1.00}{In order to illustrate the performance enhancement by the STAR-RIS enhanced SSECB design, the BSs equipped with a single transmitting antenna is communicating with one CCU and one CEU each equipped with a single receiving antenna by utilizing the power-domain NOMA techniques, which avoids the impact of the active beamforming at the BSs and detection vectors at the users.}

Without loss of generality, it is assumed that the BS located in cell1 is BS1, and the users located in cell1 treat the signals received from BS2 as interferences, whereas the users located in cell1 treat the signals received from BS1 as desired signals. The CCU and CEU located in cell1 are denoted by CCU1 and CEU1, respectively.

\textcolor[rgb]{0.00,0.00,1.00}{In this correspondence, there are two STAR-RIS arrays, namely RIS1 and RIS2, located at the intersection of two cells, each of which is employed with $L$ RIS elements with $L \ge 1$ for implementing the proposed CoMP communications.} In this Section, by properly controlling the transmission and reflection amplitude coefficients and phase shifts of each RIS element, both the desired and interference signals can be beneficially enhanced and mitigated, respectively. Note that ${{\bf{H}}^{{T}}}$ represents the transpose of the matrix $\rm \bf H$ in the rest of this correspondence.

\subsection{System Description of STAR-RIS enhanced NOMA-CoMP Networks}

Owing to the strong scattering effect of the BS-user links, the small-scale fading between BS1 and CCU1, which is the desired link, can be modelled by Rayleigh fading channel $w_{1,1,c}$, where the probability density function (PDF) is given by
\begin{equation}\label{channel matrix,Rayleigh}
{f}(x) = {e^{ - {{x}}}}.
\end{equation}

In this letter, we only focus on the impact of the path-loss, and hence the large-scale fading of the desired links for CCU1 in BS1 can be considered as Non-LoS (NLoS) link, which is given by
\begin{equation}\label{large scale fading between BS and user}
{\varepsilon_{1,1,c}} = d_{1,1,c}^{ - \alpha_1 },
\end{equation}
where $\alpha_1$ refers to the path-loss exponent of the desired link, $d_{1,1,c}$ denotes the distance of the BS1-CCU1 links.

In practice, the users located in BS1 also detect the signals from BS2, which can be modelled by
\begin{equation}\label{interference links}
I = w_{2,1,c} d_{2,1,c}^{ - {\alpha _4}},
\end{equation}
where $\alpha_4$ refers to the path-loss exponent of the interference link, $d_{2,1,c}$ refers to the distance, and $w_{2,1,c}$ refers to small-scale fading gain of the BS2-CCU1 links. Note that the small-scale fading gain of the interference links also modelled by Rayleigh fading channel. \textcolor[rgb]{0.00,0.00,1.00}{Note that the proposed network in Fig.~\ref{system_model} is a general scenario, which can be readily extended to the cellular networks with massive cells, where only received interference in~\eqref{interference links} changes.}

In the proposed CoMP networks, for simplicity, we consider that one RIS arrays and one BS are grouped, where RIS1 and RIS2 face to the selected BS1 and BS2, respectively. \textcolor[rgb]{0.00,0.00,1.00}{Since the RIS elements are located at the cell edge, the Rayleigh fading channels are adopted for modelling the NLoS of the BS1-RIS1 links as follows:
\begin{equation}\label{channel matrix,BS to LIS}
        {{\bf{H}}_1} = \left[ {\begin{array}{*{20}{c}}
                {{h_{1,R,1}}}\\
                 \vdots \\
                {{h_{1,R,L}}}
                \end{array}} \right],
\end{equation}
where ${\rm \bf H}_i$ is a $(L \times 1)$ vector containing Rayleigh fading channel gains of the BS-RIS links.}

On the one hand, the RIS elements are capable of reflecting the received signals to the users. Hence, similar to~\eqref{channel matrix,BS to LIS}, the small-scale fading vector of the reflected links, i.e., RIS1-CCU1, is defined by
\begin{equation}\label{channel matrix,LIS to w user}
 {{\rm \bf{R}}_{1,c}} = \left[ {\begin{array}{*{20}{c}}
                {{r_{1,c,1}}}& \cdots &{{r_{1,c,L}}}
                \end{array}} \right],
\end{equation}
\textcolor[rgb]{0.00,0.00,1.00}{ where ${{\rm \bf{R}}_{1,c}}$ is a $(1 \times L)$ vector representing Rician fading channel gains with fading parameter $\mathcal{K}_1$.
Correspondingly, the channel gains of the RIS-user link is computed as:
\begin{equation}\label{Rice channel gain}
r_{l,c,l} = \sqrt {\frac{{{{\cal K}_1}}}{{{{\cal K}_1} + 1}}} r_{1,c,l}^{{\rm{LoS}}} + \sqrt {\frac{1}{{{{\cal K}_1} + 1}}} r_{1,c,l}^{{\rm{NLoS}}},
\end{equation}
where $r_{1,c,l}^{{\rm{LoS}}}$ denotes the LoS components, and $r_{1,c,l}^{{\rm{NLoS}}} $ denotes the LoS and NLoS components.}

The RIS arrays are co-located at the intersection of two cells, and hence the large-scale fading channels of the BS1-RIS1 and RIS1-CCU1 can be considered as LoS links, which is defined by
\textcolor[rgb]{0.00,0.00,1.00}{ \begin{equation}\label{large scale fading reflected}
 {\varepsilon _{1,{\rm{R}},c}} = d_{1,{\rm{R}}}^{ - {\alpha _2}}d_{{\rm{R}},1,c}^{ - {\alpha _{3,c}}},
\end{equation}
where $d_{1,{\rm{R}}}$ denotes the distance of the BS1-RIS1, $d_{{\rm{R}},1,c}$ denotes the distance of the RIS1-CCU1, ${\alpha _2}$ and ${\alpha _{3,c}}$ refer to the path-loss exponents of the BS-RIS and the RIS-CCU links, respectively.}

On the other hand, the RIS elements are also capable of transmitting the received signals to the users located in the nearby cells. Hence, RIS1 transmits the signal to the users located in cell2, and the small-scale fading vector of the RIS1-CCU2 links can be defined by
\begin{equation}\label{transmitting vector}
 {{\rm \bf{T}}_{1,2,c}} = \left[ {\begin{array}{*{20}{c}}
                {{t_{1,2,c,1}}},& \cdots &,{{t_{1,2,c,L}}}
                \end{array}} \right],
\end{equation}
\textcolor[rgb]{0.00,0.00,1.00}{ where ${{\rm \bf{T}}_{1,2,c}}$ is a $(1 \times L)$ vector representing Rician fading channel gains with fading parameter $\mathcal{K}_2$, which can be written as
\begin{equation}\label{Rice channel gain_transmit link}
{t_{1,2,c,l}} = \sqrt {\frac{{{{\cal K}_2}}}{{{{\cal K}_2} + 1}}} t_{1,2,c,l}^{{\rm{LoS}}} + \sqrt {\frac{1}{{{{\cal K}_2} + 1}}} t_{1,2,c,l}^{{\rm{NLoS}}},
\end{equation}
where $t_{1,2,c,l}^{{\rm{LoS}}}$ and $t_{1,2,c,l}^{{\rm{NLoS}}}$ denote the LoS and NLoS components, respectively.}
Similarly, the large-scale fading of the BS1-RIS1-CCU2 is defined by
\textcolor[rgb]{0.00,0.00,1.00}{ \begin{equation}\label{large scale transmit }
{\varepsilon _{1,{\rm{T}},c = }}d_{1,{\rm{R}}}^{ - {\alpha _2}}d_{{\rm{R}},2,c}^{ - {\alpha _{3,c}}},
\end{equation}
where $d_{{\rm{R}},2,c}$ represents the distance of the RIS1-CCU2 links.}

\section{SSECB Beamforming Designs}

In order to simultaneously control multiple RIS elements, and to implement the proposed SSECB design, as well as based on the maturity of low-complexity channel estimation algorithms~\cite{ZhangRui_MISO_beams_discrete_2,MIMO_channel_estimation}, we assume that the global CSIs are perfectly known at the RIS controller.

The received signals at CCU1 is given by
\begin{equation}\label{received user signal}
\begin{aligned}
{y_{1,c}} & = \underbrace {\sqrt {{\varepsilon _{1,1,c}}} w_{1,1,c}^{}p + \sqrt {{\varepsilon _{1,{\rm{R}},c}}} {{\bf{R}}_{1,1,c}}{{\bf{\Phi }}_{1,{\rm{R}}}}{{\bf{H}}_1}p}_{{\rm{Desired~Signal}}} \\
&+ \underbrace {\sqrt {{\varepsilon _{2,1,c}}} w_{2,1,c}^{}p + \sqrt {{\varepsilon _{1,{\rm{T}},c}}} {{\bf{T}}_{1,2,c}}{{\bf{\Phi }}_{2,{\rm{T}}}}{{\bf{H}}_2}p}_{{\rm{Interference}}} + {N_0},
\end{aligned}
\end{equation}
\textcolor[rgb]{0.00,0.00,1.00}{ where $p$ refers to the transmit power at the BS, ${{\bf{\Phi }}_{2,{\rm{T}}}} \buildrel \Delta \over = {\rm{diag}}\left[ {{\beta _{2,{\rm{T,1}}}}{\phi _{2,{\rm{T,1}}}},{\beta _{2,{\rm{T,2}}}}{\phi _{2,{\rm{T,2}}}}, \cdots ,{\beta _{2,{\rm{T,}}L}}{\phi _{2,{\rm{T,}}L}}} \right]$ refers to both the transmitting phase shifts and transmitting amplitude coefficients of RIS2. More specifically, $\beta_{2,{\rm{T}},l}  \in \left( {0,1} \right]$ refers to the transmission amplitude coefficient of RIS element $l$ in RIS2, ${\phi _{2,{\rm{T}},l}} = \exp (j{\theta _{2,{\rm{T}},l}}), j=\sqrt{-1}, \forall l = 1,2 \cdots ,L$. ${\theta _l} \in \left[ {0,2\pi } \right)$ denotes the transmitting phase shift by RIS element $l$ in RIS2. Similarly, ${{\bf{\Phi }}_{1,{\rm{R}}}} \buildrel \Delta \over = {\rm{diag}}\left[ {{\beta _{1,{\rm{R,1}}}}{\phi _{1,{\rm{R,1}}}}, \cdots ,{\beta _{1,{\rm{R,}}L}}{\phi _{1,{\rm{R,}}L}}} \right]$ represents the reflection phase shifts and reflection amplitude coefficients of RIS1.} \textcolor[rgb]{0.00,0.00,1.00}{In this correspondence, both RIS1 and RIS2 can be perfectly controlled, where both the phase shifts and amplitude coefficients are continuous.} The additive white Gaussian noise (AWGN) is denoted by $N_0$ with variance ${\sigma ^2}$.

Based on the insights from~\cite{Yuanwei_STAR_RIS_mag,Jiaqi_STAR_RIS_lett}, there is one obvious constraint on the reflection and transmission amplitude coefficients, which is computed as:
\begin{equation}\label{amplitude constraint}
\beta _{1,{\rm{R}},l}^2{\rm{ + }}\beta _{1,{\rm{T}},l}^2 \le 1.
\end{equation}
For simplicity, it is assumed that $\beta _{1,{\rm{R}},l}^2{\rm{ + }}\beta _{1,{\rm{T}},l}^2 = 1$ in the rest of this correspondence.

To simultaneously accomplish signal enhancement and cancellation goals, the passive beamforming at both RIS1 and RIS2 must be designed jointly. We first focus our attention on cell1, and hence the design of the transmission at RIS2 is to mitigate the inter-cell interference of the users in cell1, which can be expressed as follows:

\textcolor[rgb]{0.00,0.00,1.00}{
\begin{subequations}\label{Inter-cell interference_goal}
\begin{align}
\label{eq1}
& \min \sqrt {{\varepsilon _{2,1,c}}} w_{2,1,c}^{}p + \sqrt {{\varepsilon _{2,{\rm{T}},c}}} {{\bf{T}}_{2,1,c}}{{\bf{\Phi }}_{2,{\rm{T}}}}{{\bf{H}}_2}p, \\
\label{eq2}
& \min \sqrt {{\varepsilon _{2,1,e}}} w_{2,1,e}^{}p + \sqrt {{\varepsilon _{2,{\rm{T}},e}}} {{\bf{T}}_{2,1,e}}{{\bf{\Phi }}_{2,{\rm{T}}}}{{\bf{H}}_2}p, \\
\label{eq3}
& {\rm{Subject~to}}~{\beta _{2,{\rm{T,}}l}} \le 1,\forall l{\rm{ = }}1 \cdots L,\\
\label{eq4}
&~~~~~~~~~~~~~~ {\phi _{2,{\rm{T,}}l}}\in \left[ {0,2\pi } \right), \forall l{\rm{ = }}1 \cdots L,
\end{align}
\end{subequations}
where \eqref{eq1} and~\eqref{eq2} are the interference mitigation objectives of CCU1 and CEU1, respectively. }

To achieve the ambitious objective in~\eqref{Inter-cell interference_goal}, we first generate the effective transmission channel gain of the BS2-RIS2-CCU1 and BS2-RIS2-CEU1 links as follows:
\begin{equation}\label{effective channel gain BS2-RIS2-userc}
{\overline {\bf{H}} _{2,{\rm{T}}}} = \left[ {\begin{array}{*{20}{c}}
{{h_{2,{\rm{R}},1}}{t_{2,1,c,1}}}& \cdots &{{h_{2,{\rm{R}},L}}{t_{2,1,c,L}}}\\
{{h_{2,{\rm{R}},1}}{t_{2,1,e,1}}}& \cdots &{{h_{2,{\rm{R}},L}}{t_{2,1,e,L}}}
\end{array}} \right].
\end{equation}
Then, the diagonal matrix ${{\bf{\Phi }}_{2,{\rm{T}}}}$ is transformed into a vector as follows:
\begin{equation}\label{diagnol vector of fraction of RIS2}
{\overline {\bf{\Phi }} _{2,{\rm{T}}}} = {\left[ {{\beta _{2,{\rm{T,1}}}}{\phi _{2,{\rm{T,1}}}}, \cdots ,{\beta _{2,{\rm{T,}}L}}{\phi _{2,{\rm{T,}}L}}} \right]^T}.
\end{equation}

Then, we focus on the passive beamforming design at RIS2, and the objective in~\eqref{Inter-cell interference_goal} can be rewritten as
\begin{equation}\label{refrac_rewrriten}
{\overline {\bf{H}} _{2,{\rm{T}}}}{\overline {\bf{\Phi }} _{2,{\rm{T}}}} = {{\bf I}_{2,1}},
\end{equation}
where ${{\bf I}_{2,1}}$ is a 2 $\times$ 1 vector, which represents the received inter-cell interference at CCU1 and CCU2 with ${I_ {2,1}} = \left[ {\begin{array}{*{20}{c}}
{ - \sqrt {\frac{{{\varepsilon _{2,1,c}}}}{{{\varepsilon _{2,{\rm{T}},c}}}}} w_{2,1,c}^{}}\\
{ - \sqrt {\frac{{{\varepsilon _{2,1,e}}}}{{{\varepsilon _{2,{\rm{T}},e}}}}} w_{2,1,e}^{}}
\end{array}} \right]$.
Thus, the transmission phase shifts and transmission amplitude coefficients can be readily obtained by
\begin{equation}\label{refrac_objective}
{\overline {\bf{\Phi }} _{2,{\rm{T}}}} = \overline {\bf{H}} _{2,{\rm{T}}}^{ - 1}{{\bf{I}}_{2,1}}.
\end{equation}

Similarly, RIS1 transmits the signal received from BS1 to the users located in cell2 for interference cancellation, and hence the rest of power can be reflected to the CCU1 for signal enhancement. Then, the reflection phase shifts and reflection amplitude coefficients at RIS1 can be given by
\begin{equation}\label{1st ris reflection}
{\overline {\bf{\Phi }} _{1,{\rm{R}}}} = {\left[ {\beta _{1,{\rm{R,1}}}^{}{\phi _{1,{\rm{R,1}}}}, \cdots ,\beta _{1,{\rm{R,}}L}^{}{\phi _{1,{\rm{R,}}L}}} \right]^T},
\end{equation}
where ${{\phi _{1,{\rm{R,}}l}}}$ denotes the reflection phase shifts at the $l$-th RIS element of RIS1. It is commonly assumed that the transmission and reflection phase shifts at RIS elements are independent~\cite{Yuanwei_STAR_RIS_mag,Jiaqi_STAR_RIS_lett}. Thus, the reflection of the RIS is to maximal the received power at the CCU, and the objective can be given by
\begin{equation}\label{reflection at the 1-st cell}
\begin{aligned}
&\max \sqrt {{\varepsilon _{1,1,c}}} w_{1,1,c}^{}p + \sqrt {{\varepsilon _{1,{\rm{R}},c}}} {{\bf{R}}_{1,1,c}}{{\bf{\Phi }}_{1,{\rm{Rx}}}}{{\bf{H}}_1}p,\\
& {\rm{Subject~to}}~{\beta _{1,{\rm{R,}}l}} \le 1,\forall l{\rm{ = }}1 \cdots L,\\
&~~~~~~~~~~~~~~ {\phi _{1,{\rm{R,}}l}} \in \left[ {0,2\pi } \right) , \forall l{\rm{ = }}1 \cdots L,
\end{aligned}
\end{equation}
where ${\varepsilon _{1,1,c}} = d_{1,1,c}^{ - {\alpha _1}}$ and ${\varepsilon _{1,R,c = }}d_{1,{\rm{R}}}^{ - {\alpha _2}}d_{{\rm{R}},1,c}^{ - {\alpha _3}}$ represent the large-scale fading of the BS1-CCU1 and the BS1-RIS1-CCU1 links, respectively. $w_{1,1,e}^{}$ and ${{\bf{R}}_{1,1,e}}$ denote the small-scale channel gains of BS1-CCU1 and RIS1-CCU1, respectively. ${{\bf{\Phi }}_{1,{\rm R}}}$ denotes the reflection diagonal matrix of the RIS1.

To solve the objective in~\eqref{reflection at the 1-st cell}, we then generate a reflection vector of RIS1 as follows:
\begin{equation}\label{reflection vector of first RIS}
{\overline {\bf{H}} _{1,{\rm{R,c}}}} = \left[ {\begin{array}{*{20}{c}}
{{h_{1,{\rm{R}},1}}{r_{1,{\rm{c}},1}}}& \cdots &{{h_{1,{\rm{R}},L}}{r_{1,{\rm{c}},L}}}
\end{array}} \right].
\end{equation}


Then, the reflection phase shift of RIS1 can be obtained as:
\begin{equation}\label{reflection phase shift}
\widetilde {\bf{\Phi }}_{1,{\rm{R}}}^T = \arg \left( {{{\overline {\bf{H}} }_{1,{\rm{R,c}}}}} \right) - \arg \left( {w_{1,1,c}^{}} \right),
\end{equation}
where $\arg()$ denotes the angle function. \textcolor[rgb]{0.00,0.00,1.00}{In practice, the phase shifts and amplitude coefficients are discrete due to the hardware limitations, which can be solved by a linear mapping based on an equivalent amplitude-coefficient vector as well as phase-shift vector~\cite{hou_MIMO_NOMA_SCB}.} By doing so, the received signals from the BS1-CCU1 and BS1-RIS1-CCU1 can be coherently boosted, and thus the channel gain is given by
\begin{equation}\label{effective channel gain time domain}
{\left| h \right|_{1,1,c}} = \sqrt {{\varepsilon _{1,{\rm{R}},c}}}  \sum\limits_{l = 1}^L  \left| {{h_{1,{\rm{R}},l}}{r_{1,{\rm{c}},l}}\beta _{1,{\rm{R,}}l}^{}}  \right| + \sqrt {{\varepsilon _{1,1,c}}} \left| {w_{1,1,c}^{}} \right|.
\end{equation}

\textcolor[rgb]{0.00,0.00,1.00}{ The objectives of both RIS1 and RIS2 are concluded to avoid any confusion in Table~\ref{TABLE OF objectives}.}
 \begin{table}[t!]
\small
\centering
\caption{\textcolor[rgb]{0.00,0.00,1.00}{OBJECTIVES OF RIS1 AND RIS2}}
\begin{tabular}{|c|c|}
\hline
\multirow{2}*{RIS1} & \textcolor[rgb]{0.00,0.00,1.00}{Interference mitigation for CCU2 and CEU2 }\\
\cline{2}
~     & \textcolor[rgb]{0.00,0.00,1.00}{ signal enhancement for CCU1 and CEU1}\\
\hline
\multirow{2}*{RIS2} & \textcolor[rgb]{0.00,0.00,1.00}{ Interference mitigation for CCU1 and CEU1}\\
\cline{2}
~      & \textcolor[rgb]{0.00,0.00,1.00}{ signal enhancement for CCU2 and CEU2}\\
\hline
\end{tabular}
\vspace{-0.15in}
\label{TABLE OF objectives}
\end{table}

Based on the upon joint passive beamforming designs in~\eqref{refrac_objective} and~\eqref{reflection phase shift} at two STAR-RIS arrays, the proposed SSECB design can be achieved, where the desired signals can be enhanced, while the interference signals can be cancelled simultaneously. Hence, the signal-to-interference-plus-noise-ratio (SINR) of the CEU1 can be written as
\textcolor[rgb]{0.00,0.00,1.00}{ \begin{equation}\label{SINR CEU1}
SIN{R_{1,e}} = \frac{{\left| h \right|_{1,1,e}^2p\gamma_e^2}}{{\left| h \right|_{1,1,e}^2p\gamma_c^2 + {\rho ^2}}},
\end{equation}
where ${\gamma_e^2}$ and ${\gamma_c^2}$ denote the power allocation factors of the CEU and CCU with $\gamma_e^2 + \gamma _c^2 = 1$, respectively.} Since the reflection at the RIS mainly focuses on the signal enhancement for the CCUs, the signals received at the CEUs are superimposed, and hence the channel gain of CEU1 can be given by $\left| h \right|_{1,1,e}^{} = \left| {\sqrt {{\varepsilon _{1,{\rm{R}},e}}} \sum\limits_{l = 1}^L {{h_{1,{\rm{R}},l}}{r_{1,{\rm{e}},l}}\beta _{1,{\rm{R,}}l}^{}}  + \sqrt {{\varepsilon _{1,1,e}}} w_{1,1,e}^{}} \right|$.

Based on the application of the SIC technique at the CCUs, who first detect the signal of the CEU with the following SINR:
\begin{equation}\label{CCU detects CEU}
SIN{R_{1,e \to 1,c}} = \frac{{\left| h \right|_{1,1,c}^2p\alpha _e^2}}{{\left| h \right|_{1,1,c}^2p\alpha _c^2 + {\rho ^2}}}.
\end{equation}

Then the CCUs can detect its own signal with the following signal-to-noise-ratio (SNR):
\begin{equation}\label{SNR CCU}
SN{R_{1,c}} = \frac{{\left| h \right|_{1,1,c}^2p\alpha _c^2}}{{{\rho ^2}}}.
\end{equation}

\begin{remark}\label{remark1: feasibility number of RIS}
For simplicity, it is assumed that the channels are in the strong LoS scenario, where all channel gains are one. Based on the proposed SSECB design in~\eqref{refrac_objective}, and in order to mitigate inter-cell interferences at both the CCUs and CEUs, the following constraints on the number of RIS elements need to be met:
\begin{equation}\label{feasibility analysis}
L > \max \left\{ {\sqrt {\frac{{{\varepsilon _{2,1,c}}}}{{{\varepsilon _{2,{\rm{T}},c}}}}} ,\sqrt {\frac{{{\varepsilon _{2,1,e}}}}{{{\varepsilon _{2,{\rm{T}},e}}}}} } \right\},
\end{equation}
otherwise no solution exists.
\end{remark}

Thus, the achievable rate of the CCU and CEU can be obtained by:
\begin{equation}\label{rate_CCU}
{R_{1,{\rm{c}}}} =\mathbb{E} \left\{ {{{\log }_2}\left( {1{\rm{ + }}SN{R_{1,c}}} \right)} \right\},
\end{equation}
and
\begin{equation}\label{rate_CEU}
{R_{1,{\rm{e}}}} =\mathbb{E} \left\{ {{{\log }_2}\left( {1{\rm{ + }}SIN{R_{1,e}}} \right)} \right\},
\end{equation}
where $\mathbb{E}\{\}$ represents the expectation function.

\section{Numerical Results}

In this section, numerical results for evaluating the proposed SSECB design in the STAR-RIS enhanced NOMA-CoMP networks are presented. Note that in this paper, the transmission bandwidth of the proposed network is set to $B=1$ MHz, and the power of the AWGN is set to ${\rho ^2} =  - 174 + 10{\rm{lo}}{{\rm{g}}_{10}}(B)$ dBm. \textcolor[rgb]{0.00,0.00,1.00}{ The paired users share the power with the power allocation factors $\gamma_e^2=0.6$ and $\gamma _c^2=0.4$ on the basis of NOMA protocol, which is a common parameter setting in previous contributions with fixed-power allocation NOMA~\cite{Hou_Single_UAV}. } \textcolor[rgb]{0.00,0.00,1.00}{Note that more sophisticated optimization on the power allocation factors is capable of further enhancing the network's performance, however, the power allocation factor optimization is beyond the scope of this correspondence. }

Some detailed parameters are illustrated in Table~\ref{TABLE OF PARAMETERS}.
\begin{table}[t!]
\small
\centering
\caption{\small{SIMULATION PARAMETERS}}
\begin{tabular}{|c|c|}
\hline
The distances of the BS1-CCU1 and BS2-CCU2 &30 m\\
\hline
The distances of the BS1-CEU1 and BS2-CEU2 &60 m\\
\hline
The distances of the BS1-RIS1 and BS2-RIS2 &70 m\\
\hline
The distances of the RIS1-CEU1 and RIS2-CEU2 &15 m\\
\hline
The distances of the RIS1-CCU1 and RIS2-CCU2 &50 m\\
\hline
The distances of the BS1-CCU2 and BS2-CCU1 &120 m\\
\hline
The distances of the BS1-CEU2 and BS2-CEU1 &90 m\\
\hline
The path-loss exponent of BS-user links  & $\alpha_1=3$ \\
\hline
The path-loss exponent of BS-RIS links  & \textcolor[rgb]{0.00,0.00,1.00}{ $\alpha_2=3$} \\
\hline
The path-loss exponent of RIS-CEU links  &\textcolor[rgb]{0.00,0.00,1.00}{ $\alpha_{3,e}=2.4$} \\
\hline
The path-loss exponent of RIS-CCU links  &\textcolor[rgb]{0.00,0.00,1.00}{ $\alpha_{3,c}=2.7$ }\\
\hline
The path-loss exponent of interference links  & $\alpha_4=3.5$ \\
\hline
The fading parameters& ${\cal{K}}_1=2$ ${\cal{K}}_2=3$ \\
\hline
\end{tabular}
\vspace{-0.15in}
\label{TABLE OF PARAMETERS}
\end{table}

\begin{figure}[t!]
\centering
\includegraphics[width =3in]{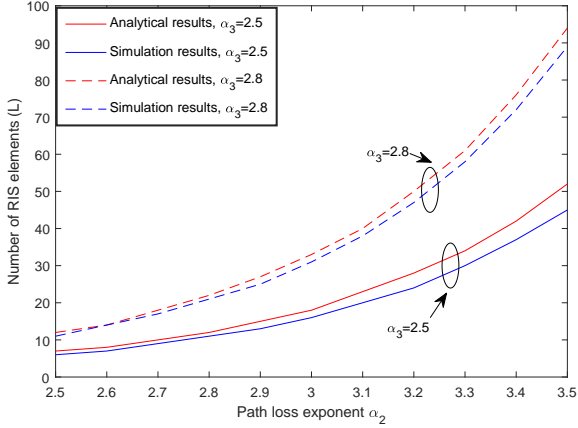}
\caption{\textcolor[rgb]{0.00,0.00,1.00}{ Minimal required number of RIS elements for the proposed SSECB design in the STAR-RIS enhanced NOMA-CoMP networks, where the small-scale channel gains are normalized to all one, and $\alpha_{3,c}=\alpha_{3,e}=\alpha_3$ for simplicity.}}
\label{Minimum_number of RIS_fig2}
\vspace{-0.15in}
\end{figure}

\emph{1) Minimal Required Number of RIS elements:} The minimal required number of RIS elements of each RIS array for implementing the proposed SSECB design is evaluated in Fig.~\ref{Minimum_number of RIS_fig2}. It is observed that with the increases of the path loss exponent, the minimal required number of RIS elements increases, which indicates that the path-loss exponent of the BS-RIS and RIS-user links are required to be higher than that of the interference links of the proposed SSECB design. Furthermore, note that in the simulation results, it is also assumed that the small-scale channel gains are also normalized to one for simplicity, and hence, the number of RIS elements should be greater than the minimal number of RIS elements in practice.

\begin{figure}[t!]
\centering
\includegraphics[width =3in]{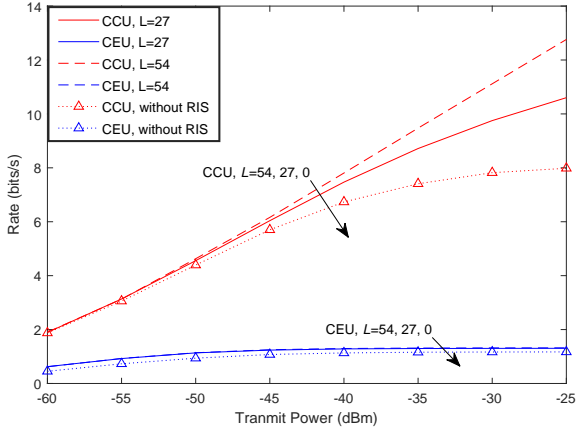}
\caption{\textcolor[rgb]{0.00,0.00,1.00}{ The achievable rate of both the CCU and CEU versus the transmit power with different number of RIS elements. The minimal number of RIS element is obtained to $L=27$ for the selected parameters.}}
\label{STAR_RIS_with diff number of RIS}
\vspace{-0.15in}
\end{figure}

\emph{2) Impact of the Number of RIS Elements:} The achievable rate of both the CCU and CEU in the proposed SSECB design are evaluated in Fig.~\ref{STAR_RIS_with diff number of RIS}. The number of RIS elements is obtained by~\eqref{feasibility analysis}, where $L=27$ in the case of the selected parameter setting.

By comparing the red solid curve and dashed curve, we can see that in the case of $L=54$, which is two times of the minimal required number of RIS elements, the high-SNR slope of the achievable rate is one, which illustrates that the interferences received from nearby cells are mitigated perfectly. One can also observe that the achievable rate ceilings occur in the case of $L=27$, which is due to the fact that the number of RIS elements is not enough for interference cancellation, and hence the inter-cell interference residue exists. Observe that the performance gap between the proposed SSECB design and the case without RIS increases, which also show the superiority of the proposed SSECB design.

\begin{figure}[t!]
\centering
\includegraphics[width =3in]{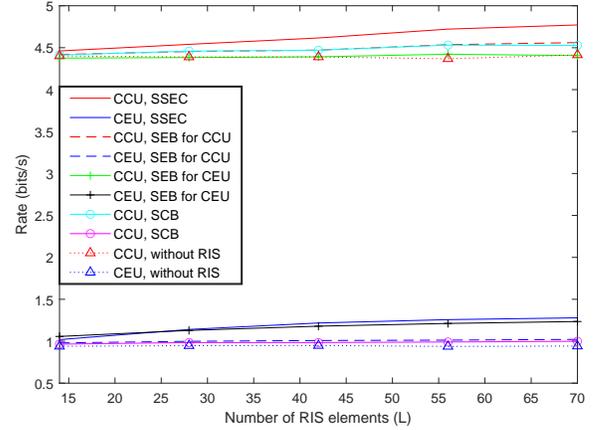}
\caption{The achievable rate of both the CCU and CEU versus the number of RIS elements with different passive beamforming designs. The transmit power of the BS is set to $p=-50$ dBm.}
\label{different passive beamformign deisngs fig4}
\vspace{-0.1in}
\end{figure}

\emph{3) Impact of Different Passive Beamforming Designs:} The achievable rate of the paired NOMA users with different passive beamforming at RIS elements are evaluated in Fig.~\ref{different passive beamformign deisngs fig4}. By aligning the reflected signals and desired signal perfect coherent at the CCU or CEU, the dashed, green and black curves are obtained for comparing the performance of a pair of SEB designs. We can see that the achievable rate of the proposed SSECB design is higher than other benchmark schemes, which illustrate that the proposed SSECB design is capable of outperforming the other SEB and SEC designs.

\section{Conclusions}
In this letter, firstly, previous contributions related to the RIS-enhanced SEB and SCB designs are presented. Then, a novel STAR-RIS for the proposed SSECB design was utilized. We designed a joint passive beamforming weight, where the transmission and reflection of RIS elements are used for inter-cell interference cancellation and signal enhancement, respectively, to provide a practical design of the STAR-RIS enhanced NOMA-CoMP networks. Compared to the previous SEB and SCB designs, our proposed STAR-RIS enhanced SSECB design has superior performance on achievable rate. Further research should be undertaken to add more TA and RA at the BS and users by jointly designing both the active beamforming, passive beamforming, and detection vectors. \textcolor[rgb]{0.00,0.00,1.00}{ Another future direction is to combine stochastic geometry tools for evaluation the impact of location randomness of users~\cite{9110835}.}

\bibliographystyle{IEEEtran}
\bibliography{IEEEabrv,STAR_RIS_CoMP_SSEC_letter}

\end{document}